\documentstyle[12pt]{article}
\input{psfig.sty}
\begin{document}

\begin{titlepage}

\begin{flushright}
RAL-TR-98-021
\end{flushright}

\baselineskip 24pt

\begin{center}

{\Large {\bf Neutrino Oscillations in the Dualized Standard Model}}\\

\vspace{.5cm}

\baselineskip 14pt
{\large Jos\'e BORDES}\\
bordes\,@\,evalvx.ific.uv.es\\
{\it Dept. Fisica Teorica, Univ. de Valencia,\\
  c. Dr. Moliner 50, E-46100 Burjassot (Valencia), Spain}\\
\vspace{.2cm}
{\large CHAN Hong-Mo}\\
chanhm\,@\,v2.rl.ac.uk\\
{\it Rutherford Appleton Laboratory,\\
  Chilton, Didcot, Oxon, OX11 0QX, United Kingdom}\\
\vspace{.2cm}
{\large Jakov PFAUDLER}\\
jakov\,@\,thphys.ox.ac.uk\\
{\it Dept. of Physics, Theoretical Physics, University of Oxford,\\
  1 Keble Road, Oxford, OX1 3NP, United Kingdom}\\
\vspace{.2cm}
{\large TSOU Sheung Tsun}\\
tsou\,@\,maths.ox.ac.uk\\
{\it Mathematical Institute, University of Oxford,\\
  24-29 St. Giles', Oxford, OX1 3LB, United Kingdom}
\end{center}

\vspace{.3cm}

\begin{abstract}

A method developed from the Dualized Standard Model for calculating the
quark CKM matrix and masses is applied to the parallel problem in neutrino 
oscillations.  Taking the parameters determined from quarks and the masses
of two neutrinos: $m_3^2 \sim 10^{-2} - 10^{-3} {\rm eV}^2$ suggested by
atmospheric neutrino data, and $m_2^2 \sim 10^{-10} {\rm eV}^2$ suggested 
by the long wave-length oscillation (LWO) solution of the solar neutrino
problem, one obtains from a parameter-free calculation all the mixing angles 
in reasonable agreement with existing experiment.  However, the scheme is
found not to accommodate comfortably the mass values $m_2^2 \sim 10^{-5}
{\rm eV}^2$ suggested by the MSW solution for solar neutrinos.

\end{abstract}

\end{titlepage}

\clearpage

Experiments of recent years have accumulated an increasing amount of quite 
convincing evidence for the existence of neutrino oscillations which is 
beginning seriously to constrain the theoretical models invented for their 
explanation \cite{review}.  The problem thus offers on the one hand a possible 
window into a region of physics which is so far unexplored and, on the 
other, a challenge and a valuable testing ground for any theory which 
attempts to understand the many intriguing features of the Standard Model 
as we know it today.  In particular, it would be interesting to ask whether 
the mass and mixing patterns we see in the quarks and the charged leptons 
are reflected in some way in the neutrinos, and if so why it is that the 
neutrinos should appear nevertheless to be so very different, for example 
in the extreme smallness of their mass and in the apparent absence of their 
right-handed partners. 

Now we have recently suggested a scheme called the Dualized Standard Model 
\cite{Chantsou} which purports to have explained with some success the mass 
and mixing patterns of the quarks and the masses of the charged leptons
\cite{OurCKM}.  Thus, it would seem incumbent upon us to make an attempt 
also at explaining neutrino oscillations with the same methodology.  The 
purpose of the present article is to make a start in doing so.

The Dualized Standard Model (DSM) is based on a recent theoretical result that 
nonabelian Yang-Mills theory has an analogue of the electric-magnetic duality 
of the abelian theory via a generalized dual transform \cite{Chanftsou}.  This
implies in particular that in addition to the (electric) $SU(3)$ colour 
symmetry the Standard Model has also a dual (magnetic) $\widetilde{SU}(3)$
symmetry.  The $SU(3)$ colour symmetry being, as we know, in the confined phase,
it then follows from a well-known result of 't~Hooft's \cite{thooft} that the
$\widetilde{SU}(3)$ dual colour symmetry is in the Higgs phase and broken.
Fermions occuring in the triplet representation of $\widetilde{SU}(3)$ (which 
are actually monopoles of colour) would then carry a broken dual colour index 
which would be similar in appearance to the generation index.  If we choose 
to identify the two indices, as we did in the DSM scheme \cite{Chantsou}, 
then it follows that there are 3 and only 3 generations, a fact which seems 
strongly supported by present experiment.

The scheme further predicts that, at the tree-level, only the highest
generation fermion has a mass and that the CKM mixing matrix between the
$U$-type and $D$-type quarks is the identity, which is already not a bad 
zeroth order approximation to the experimental picture.  Moreover, it 
goes on to predict that loop-corrections will lift this degeneracy, and 
even suggests a method whereby such loop-corrections can be perturbatively 
calculated \cite{Chantsou}.  A calculation to 1-loop order has already been 
performed, which shows that with only a small number of parameters, one gets a 
good fit to the experimental CKM matrix and sensible values also for the quarks
and charged leptons masses \cite{OurCKM}.  It seems therefore natural, perhaps 
even unavoidable, to ask whether the same procedure would apply also to 
neutrinos.

\begin{figure}[htb]
\vspace{-5cm}
\centerline{\psfig{figure=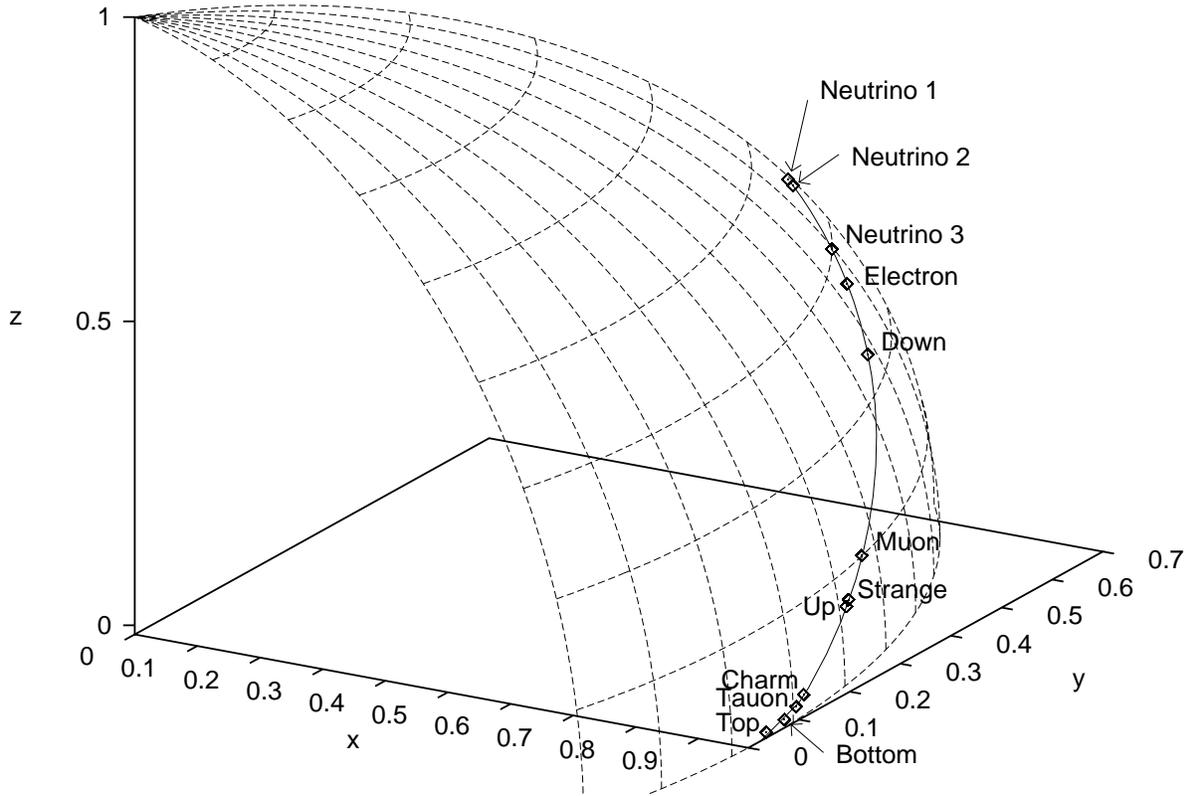,width=0.9\textwidth}}
\vspace{-1cm}
\caption{The trajectory traced out by $(x', y',z')$.}
\label{runtraj}
\end{figure}

To set up the enquiry, let us first recall that in the DSM scheme, the 
fermion mass matrix remains factorizable to all orders, namely that:
\begin{equation}
M' = M_T \left( \begin{array}{c} x' \\ y' \\z' \end{array} \right) (x',y',z'),
\label{fermasmat}
\end{equation}
where $M_T$ is the mass of the highest generation.  Everything we need to 
know for calculating the CKM mixings and the fermion masses is encoded in 
the vector $(x',y',z')$, which for the questions we ask here we may take 
to be normalized thus: 
\begin{equation}
x'^2 + y'^2 + z'^2 = 1.
\label{normxyz}
\end{equation}
This vector $(x',y',z')$ rotates with the energy scale, and thus traces out a
trajectory on the unit sphere, starting from near a fixed point $(1,0,0)$ 
at high energy scales to near another fixed point $\frac{1}{\sqrt{3}}(1,1,1)$
at low energies, as illustrated in Figure \ref{runtraj}.  The actual 
trajectory it traces out depends on $(x,y,z)$, the tree-level values of 
$(x',y',z')$ which are also the vacuum expectation values of the Higgs 
fields which break the dual colour symmetry, and also weakly on the 
strength $\rho$ of the Yukawa coupling of these Higgs fields to the 
fermion under consideration.  The vev's $(x,y,z)$ are common to all 
fermion-types but, as far as we understand it at present, the $\rho$'s can 
in principle be different for different fermion-types.  However, for some 
yet unknown reason, the 3 values obtained by fitting the quark CKM matrix 
and the two higher generation quark and charged lepton masses turned out 
to be equal to a high accuracy, so much so that we suspect that there is 
a hidden symmetry in the problem which we have not yet understood.  The 
result in practical terms is that all 3 fermion-types ($U, D, L$) run on 
the same trajectory, and they differ only in the positions where the actual 
physical states of each type are located on that trajectory.  The trajectory
shown in Figure \ref{runtraj} is in fact the one determined in \cite{OurCKM}
by fitting the quark CKM matrix and masses.  Also shown are the locations
of the various quark and charged lepton states on the trajectory obtained 
in that calculation.  In this scenario, the masses and state vectors of 
all three generations of each fermion type are obtained once the mass $M_T$ 
of the highest generation appearing in (\ref{fermasmat}) is specified.  For 
more details, the reader is referred to our earlier paper \cite{OurCKM}.

Let us turn now to the problem of neutrinos.  Since neutrinos seem to exist 
also in three generations \cite{lep} which in the DSM scheme would be 
identified with dual colour, it would appear that nothing is changed compared 
with the other three fermion-types as far as their Dirac mass matrix is 
concerned.  Hence, once given the mass $M_3=M_T$ of the highest generation
and assuming the same $\rho$ as for the other fermions, our prescription 
will allow us to calculate the Dirac masses $M_2$ and $M_1$ of the two 
lower generations as well as the state vectors of all three generations.
And since the state vectors of the charged leptons are already known from our 
earlier work \cite{OurCKM}, one can then calculate also the leptonic CKM matrix
and hence the mixing angles appearing in neutrino oscillations.  However,
the Dirac masses $M_i$ specified above are not yet the physical masses
of the neutrino states, for, as is well known, right-handed neutrinos
can have Majorana masses, so that for each generation $i$ one has yet 
to diagonalize a $2 \times 2$ submatrix of the form:
\begin{equation}
{\bf M}_i = \left( \begin{array}{cc} 0 & M_i \\ M_i & B \end{array} \right),
\label{bfmasmat}
\end{equation}
giving for the physical masses of the neutrinos:
\begin{equation}
m_i = M_i^2/B,
\label{seesaw}
\end{equation}
where for the DSM as formulated in \cite{Chantsou} $B$ has to be the same for 
all $i$ for consistency.  For $B$ large, this way of determining the physical
masses of neutrinos is the famous see-saw mechanism \cite{seesawref} which 
can give very small physical masses $m_i$ for the neutrinos with not too 
small Dirac masses $M_i$.  The parameter $B$, which can be interpreted as the 
mass of the yet undiscovered neutrinos with a large right-handed component 
(henceforth referred to as `right-handed neutrinos' in short), is unknown, 
so that in contrast to the other fermion-types depending on only one mass 
scale, the neutrino calculation involves two mass scales which have still to be
specified.  This we can do once we know the masses of any two of the neutrinos.

We turn then to experiment to see whether we can find enough information to
determine two of the neutrino masses.  Attempts at direct measurements of
neutrino masses have yielded up to now the following sort of upper limits:
\cite{databook}
\begin{equation}
m_{\nu_\tau} < 24 {\rm MeV}, \; m_{\nu_\mu} < 0.17 {\rm MeV}, \; 
   m_{\nu_e} < 10 {\rm eV},
\label{masslim}
\end{equation}
which one suspects to be rather weak.  On the other hand, information from 
other sources, such as the depletion effects of solar and atmospheric 
neutrinos when interpreted as being due to neutrino oscillations is much more
stringent.  For example, to explain the solar neutrino puzzle as neutrino
oscillations, we are offered two solutions: (i) the so-called long-wave
length oscillation solution (LWO) which corresponds to oscillations over 
distance scales of the order of the radius of the earth's orbit and requires 
$\Delta m^2_{12} \sim 10^{-10} {\rm eV}^2$ \cite{Bargphil,Krascov}, and 
(ii) the Mikheyev-Smirnov-Wolfenstein (MSW) solution \cite{MSW,MSWrev} which 
corresponds to oscillations over distance scales of the order of the sun's 
radius and requires $\Delta m^2_{12} \sim 10^{-5} {\rm eV}^2$.  On the 
other hand, to explain the atmospheric neutrino anomaly observed by the 
Kamiokande \cite{Kamioka}, IMB \cite{IMB} and Soudan experiments 
\cite{Soudan}, $\Delta m^2_{23} \sim 10^{-2} - 10^{-3} {\rm eV}^2$ is 
required \cite{Giunkimno}.  Here $\Delta m^2_{ij}$ represents the mass 
squared difference between the $i$th and $j$th generation neutrinos. 

As far as we are concerned within the framework of the DSM, where masses 
for the lower two generations of fermions are obtained by the `leakage'
mechanism of a rotating mass matrix as explained in \cite{OurCKM}, the Dirac
masses $M_i$ of the neutrinos, like other fermions, have to be hierarchical,
implying thus by (\ref{seesaw}) $m_1 \ll m_2 \ll m_3$.  Hence, from the above 
estimates for mass differences, we conclude that we should put for the mass of 
the heaviest neutrino $m^2_3 \sim 10^{-2} - 10^{-3}$ eV$^2$, and for the mass 
of the second heaviest neutrino either (i) $m^2_2 \sim 10^{-10}$ eV$^2$ (LWO),
or (ii) $m_2^2 \sim 10^{-5}$ eV$^2$ (MSW).  We have thus the two mass scales
we seek as input information for proceeding with our calculation, which, 
apart from the two masses here determined from experiment, will be parameter 
free, and will give us as predictions all the mixing angles in the leptonic 
CKM matrix, as well as the masses of the lightest neutrino and the right-handed 
neutrinos.

The calculation follows along exactly the same lines as for the quarks
and charged leptons given in \cite{OurCKM} and uses essentially the same
numerical programs.  Starting with some assumed value for $M_3$ the Dirac
mass, one runs the scale down until it becomes equal to the Dirac mass 
$M_2$ of the second generation obtained by the `leakage' mechanism from
the rotating mass matrix.  From these values of $M_3$ and $M_2$, one can
then evaluate the corresponding values of $m_2/m_3 = M_2^2/M_3^2$, the
ratio of the actual masses of the two heaviest neutrinos as given by 
the see-saw mechanism (\ref{seesaw}).  This value of the mass ratio, of
course, need not agree with either of the values obtained from the estimates
given in the above paragraph deduced from experiment.  One then varies the
assumed value for $M_3$ until the output ratio for $m_2/m_3$ agrees with
the value favoured by experiment.  For each assumed value of $M_3$, our 
program automatically calculates the corresponding value for $m_1$, the 
mass of the lightest neutrino, and the predicted mass of the right-handed 
neutrino $B$.  Furthermore, the orientation in dual colour space of the 
state vectors for all three generations of neutrinos are also given.  Hence, 
combining this with the result previously obtained in \cite{OurCKM} for 
the state vectors of the charged leptons, the whole leptonic CKM matrix 
can be evaluated.

Consider first the case corresponding to the long wave-length (LWO) solution
(i) for the solar neutrino puzzle, namely $m_2^2 \sim 10^{-10}$ eV$^2$.
We obtain for input values for $M_3 = 2.0, 3.7$ MeV respectively the 
following results:
\begin{equation}
m_3^2 = 10^{-2} {\rm eV}^2, \; m_1 = 5 \times 10^{-17} {\rm eV}, \; 
   B = 40 {\rm TeV};
\label{numasses1}
\end{equation}
\begin{equation}
|CKM|_{lepton} = \left( \begin{array}{ccc} 0.9671 & 0.2417 & 0.0795 \\
                                  0.2277 & 0.6823 & 0.6947 \\
                                  0.1136 & 0.6899 & 0.7149 \end{array} \right),
\label{lepCKM1}
\end{equation}
and:
\begin{equation}
m_3^2 = 10^{-3} {\rm eV}^2, \; m_1 = 10^{-15} {\rm eV}, \; B = 430 {\rm TeV};
\label{numasses2}
\end{equation}
\begin{equation}
|CKM|_{lepton} = \left( \begin{array}{ccc} 0.9694 & 0.2355 & 0.0700 \\
                                  0.2215 & 0.7142 & 0.6640 \\
                                  0.1063 & 0.6591 & 0.7445 \end{array} \right).
\label{lepCKM2}
\end{equation}
We notice first that it is possible to obtain values of $m_3$ within the 
range required by the atmospheric neutrino experiments.  Secondly, we 
note that the mass obtained for the lightest neutrino is extremely small.  
This is because the value of $(x',y',z')$ for this neutrino, as seen in 
Figure \ref{runtraj}, is already getting very close to the fixed point 
$\frac{1}{\sqrt{3}}(1,1,1)$ so that the leakage mechanism hardly operates 
and thus gives it very little mass.  Needless to say, our calculation in 
this mass region is far from reliable and gives at best just a rough order 
of magnitude.  The estimate for the mass of the right-handed neutrino $B$ 
depends only on the two heaviest neutrinos and should be more reliable.
Interestingly, its value turns out to be of the same order as the value
of the vev's of the Higgs fields responsible for breaking the dual colour
or generation symmetry as estimated from the experimental bounds on 
the $K^0- {\bar K}_0$ mass difference and on flavour-changing neutral 
current decays \cite{OurFCNC}.  Notice that our estimate for $B$ is 
considerably lower than what is usually assumed, in grand unified theories, 
for example \cite{Ross}.  The reason is that one usually uses a Dirac mass for 
the neutrino similar to that for the charged lepton, namely for the highest
generation a mass of around 1 GeV.  On the other hand, for our calculation 
here we want for the Dirac mass $M_3$ a value of only a few MeV, and for
fixed $m_3$, $B$ according to (\ref{seesaw}) is proportional to $M_3^2$.  
That neutrinos and charged leptons can have widely different Dirac masses 
one need not find disturbing if one recalls that even for the quarks, Dirac 
masses between the $U$- and the $D$-types differ by as much as a factor 50, 
as witnessed by the masses of the $t$ and the $b$.  The fact that our
estimate for $B$ is of order $10^3$ TeV means that the implied limits for
neutrinoless double beta decay and for neutrino-antineutrino
oscillations, while compatible with existing experimental bounds,
may be much more accessible than previously anticipated.  A detailed
analysis of the experimental situation has not, however, been done and
is beyond the scope of the present work.

In this paper, we focus on the mixing angles in the leptonic CKM matrix 
for comparison with existing data.  To order 1-loop corrections in the DSM 
scheme, the CKM matrix whether for quarks or leptons is real \cite{OurCKM}, 
so that there are only three independent parameters to consider, which we 
can take to be the three elements in the upper right corner of the matrix, 
namely $U_{\mu 3}, U_{e 3}$, and $U_{e 2}$.  We shall examine each in turn.  

Consider first the element $U_{\mu 3}$ which plays the central role in 
atmospheric neutrinos, where, to explain the muon puzzle, one needs a 
sizeable value for $U_{\mu 3}$.  Indeed, according to the recent analysis
in \cite{Giunkimno}, for example, $|U_{\mu 3}|$ has to have a value roughly 
between 0.45 and 0.85 to explain the Kamiokande data \cite{Kamioka}.  One 
notes that the values we obtained in (\ref{lepCKM1}) and (\ref{lepCKM2})
fall right in the middle of the permitted range.

Next, consider the element $U_{e 3}$.  Its value is constrained not only 
by atmospheric neutrino data but also by terrestial experiments such as
Bugey \cite{Bugey} and CHOOZ \cite{Chooz}.  The absence of any observed 
effects in the latter type of experiments puts an upper bound on $|U_{e 3}|$ 
of around 0.15 at, for example, $m_3^2 \sim 10^{-2} {\rm eV}^2$.  Again, one 
notes that the value we calculated, as quoted in (\ref{lepCKM1}) and 
(\ref{lepCKM2}) above, satisfy this bound.

\begin{figure}[htb]
\vspace{-3cm}
\centerline{\psfig{figure=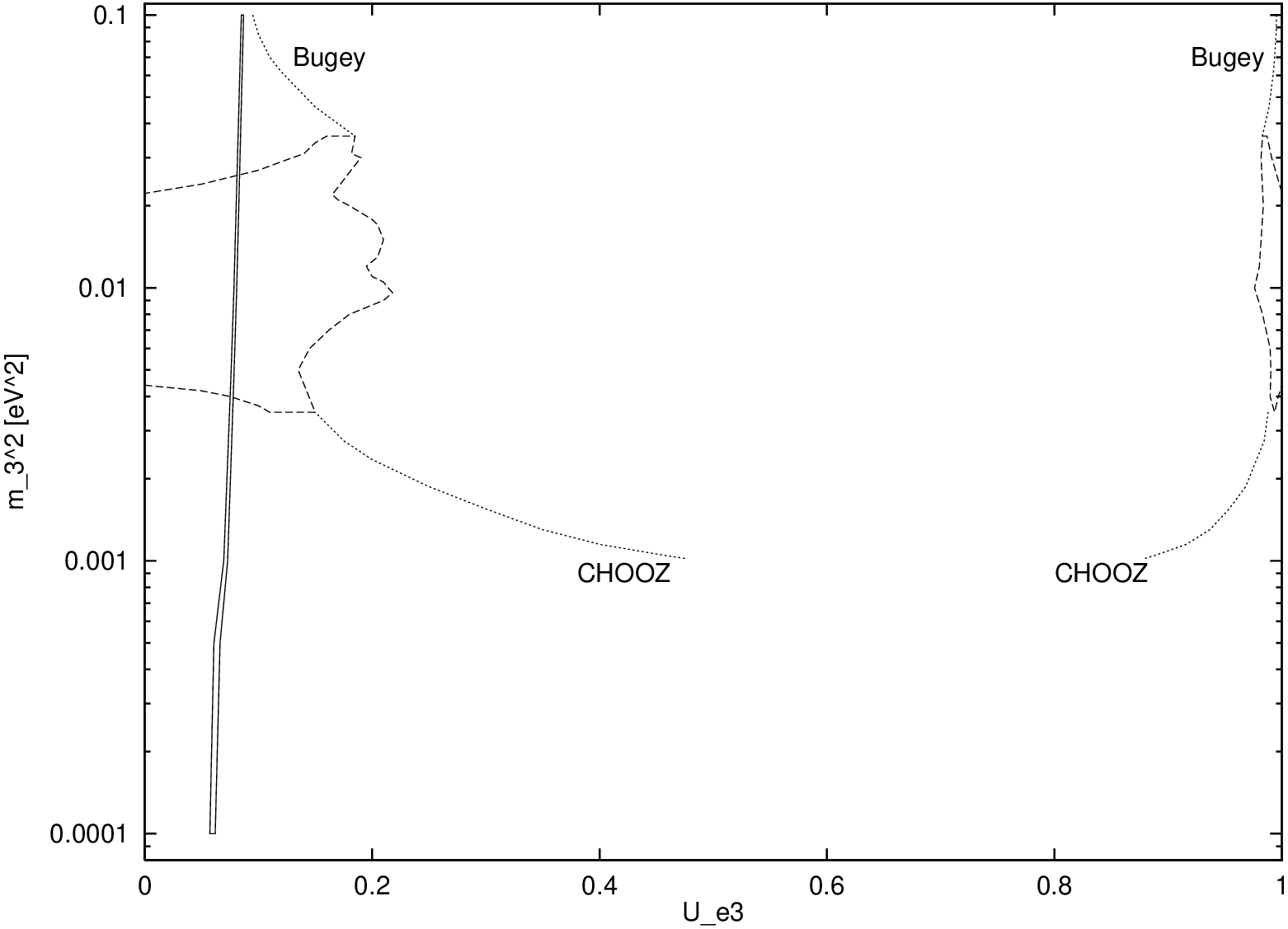,width=0.75\textwidth}}
\vspace{0cm}  
\caption{90\% CL limits on the CKM element $U_{e 3}$ compared with the 
result from our calculation.}
\label{Ue3}
\end{figure}

\begin{figure}[htb]
\vspace{-3cm}
\centerline{\psfig{figure=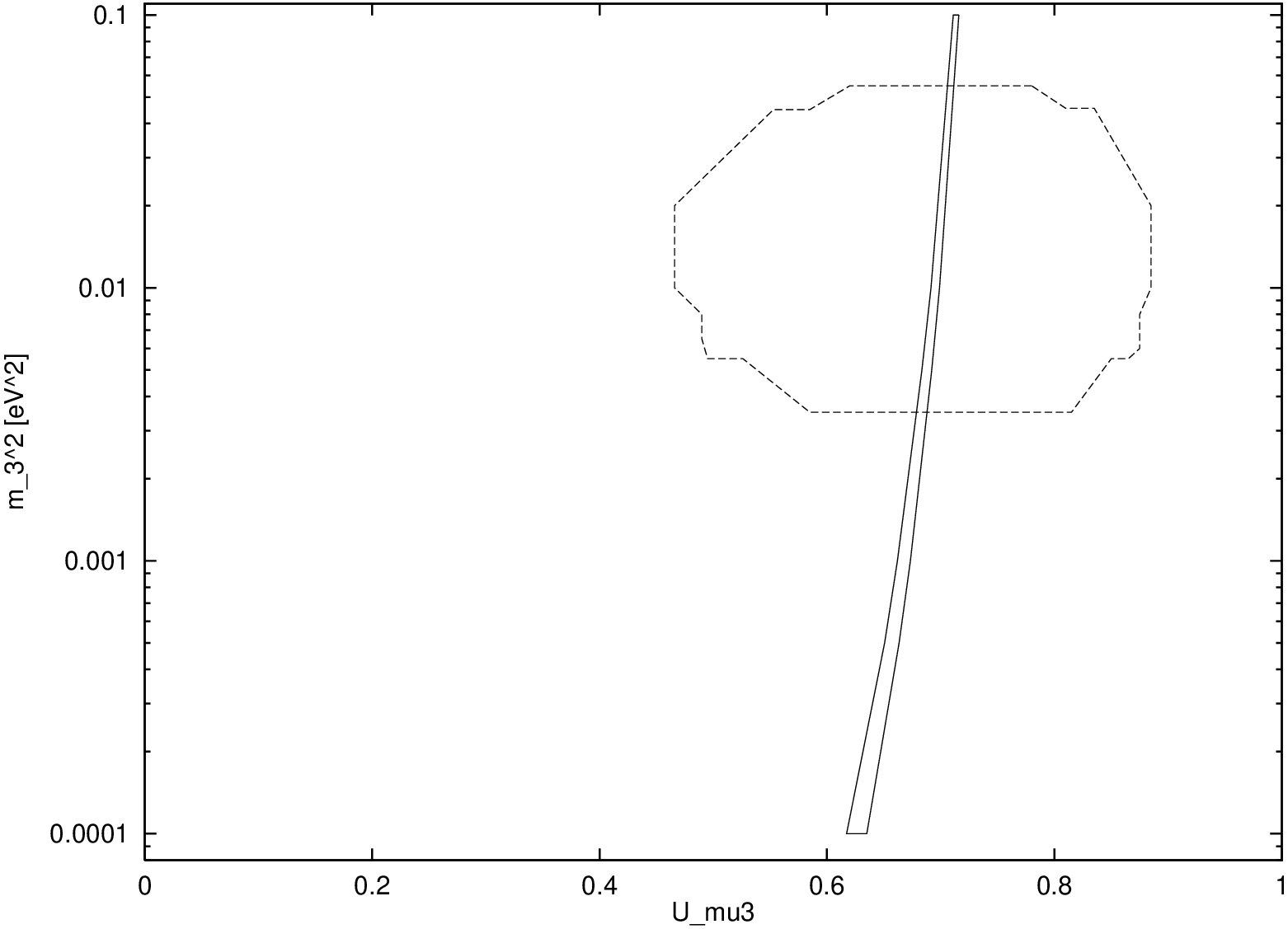,width=0.75\textwidth}}
\vspace{0cm}  
\caption{90\% CL limits on the CKM element $U_{\mu 3}$ compared with the
result from our calculation.}
\label{Umu3}
\end{figure}

\begin{figure}[htb]
\vspace{-3cm}
\centerline{\psfig{figure=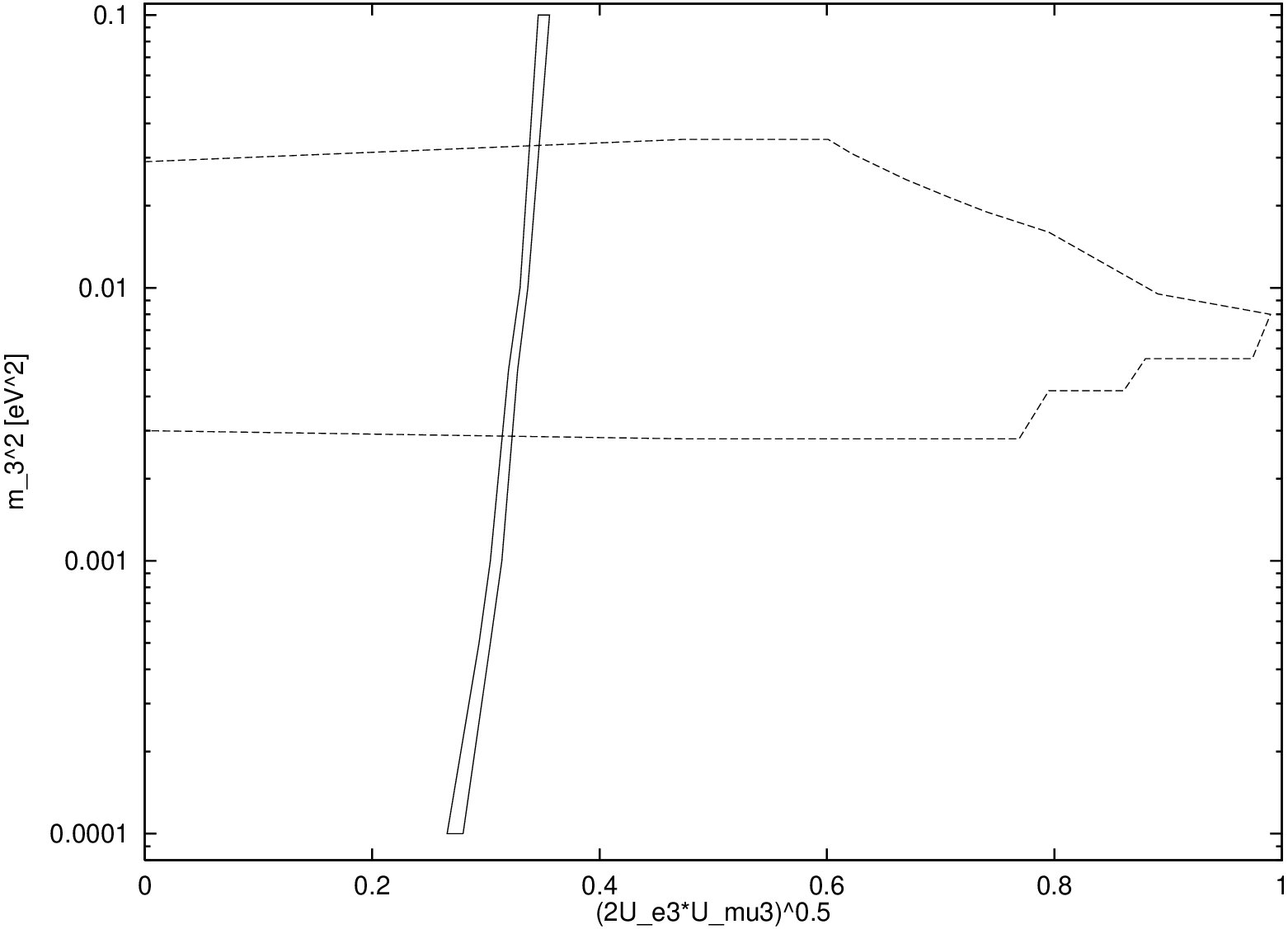,width=0.75\textwidth}}
\vspace{0cm} 
\caption{90\% CL limits on the quantity $(2 U_{e 3} U_{\mu 3})^{1/2}$ compared
with the result from our calculation.}
\label{rtUe3Umu3}
\end{figure}

\begin{figure}[htb]
\vspace{-3cm}
\centerline{\psfig{figure=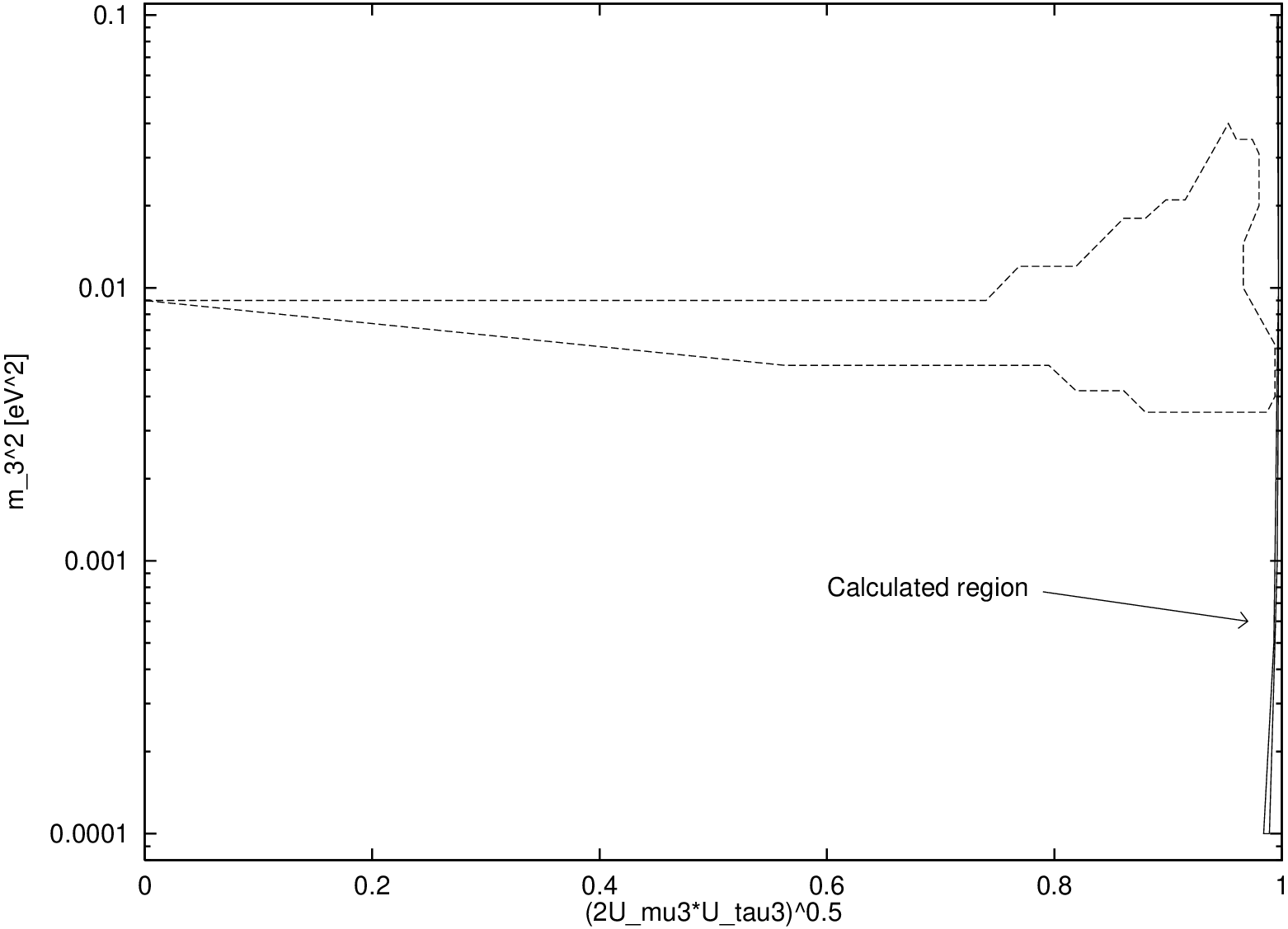,width=0.75\textwidth}}
\vspace{0cm}  
\caption{90\% CL limits on the quantity $(2 U_{\mu 3} U_{\tau 3})^{1/2}$
compared with the result from our calculation.}
\label{rtUmu3Utau3}
\end{figure}

In Figure \ref{Ue3} and \ref{Umu3}, we reproduce in terms of the CKM matrix 
elements the individual 90\% CL limits on $U_{e 3}$ and $U_{\mu 3}$ obtained 
by \cite{Giunkimno} with the data from the Kamiokande, Bugey and CHOOZ
experiments and compare them with the result of our calculation for a range
of $m_3$ values.  Also shown in Figures \ref{rtUe3Umu3} and \ref{rtUmu3Utau3}
are the correlated bounds on these elements quoted from the same source.
The width of our curves represents the range of values calculated from
$m_2^2$ values lying within the admissible range $5 \times 10^{-6}$ to 
$1.1 \times 10^{-5}$ eV$^2$ obtained from the analysis of solar neutrino 
data by \cite{Bargphil,Krascov}.  One notes first that the calculated
values of the mixing parameters are quite insensitive both to the input 
value of $m_2^2$ and to the value of $m_3^2$.  Secondly, one sees that 
in each case our curve lies comfortably within the experimental limits 
for all reasonable values of $m_3^2$ except for Figure \ref{rtUmu3Utau3}
where it passes near the edge.  It seems thus that the agreement of
our calculated $U_{\mu 3}$ and $U_{e 3}$ with experiment is rather good.
Notice that the analysis in \cite{Giunkimno} did not take into account the 
new SuperKamiokande \cite{superk} data which is thought to lower the estimate 
for $m_3^2$ to around $10^{-3}$ eV$^2$.  For this reason we have given the 
result of our calculation also for $m_3^2$ values outside the quoted limits 
from \cite{Giunkimno} in anticipation of a comparison with future analyses of
the SuperKamiokande data.
  
For the remaining CKM element $U_{e 2}$ which enters mainly in the solar 
neutrino problem, we cannot as yet make a clear comparison with experiment.  
Detailed analyses of the solar neutrino data for the long wave-length (LWO)
oscillation scenario have, as far as we know, been performed only for two
flavours \cite{Bargphil,Krascov}.  If we compare our calculated $U_{e 2}$
with these two-flavour analyses, then our values for the mixing angle of 
around 14$^o$ lie outside the bound obtained of $> 27^o$.  However, a full
three-flavoured analysis of the solar neutrino data where accounts are
taken of 
both long wave length and MSW oscillations has to be done before a meaningful 
comparison can be made, since for  $m_3^2$ approaching $2 \times 
10^{-4} {\rm eV}^2$, a value possibly compatible with SuperKamiokande 
\cite{superk}, an adiabatic MSW transition
$e \rightarrow 3$ may lead to significant depletion.  Even if it turns out 
that the quoted estimate from the two-flavoured analyses is not appreciably 
affected with 3 flavours taken into account so that the discrepancy with 
our calculation remains, one should perhaps still not be too disappointed, 
given that the value is obtained from a parameter-free calculation in which 
fairly crude approximations have been made, and that in our scheme
$U_{e2}$ is particularly sensitive to details \cite{curves}.

It is instructive to compare the lepton CKM matrix (\ref{lepCKM1}) or
(\ref{lepCKM2}) obtained above with the quark CKM matrix calculated in 
\cite{OurCKM} by the same method with a common set of parameters:
\begin{equation}
|CKM|_{quark} = \left( \begin{array}{ccc} 0.9755 & 0.2199 & 0.0044 \\
                                  0.2195 & 0.9746 & 0.0452 \\
                                  0.0143 & 0.0431 & 0.9990 \end{array} \right),
\label{quaCKM}
\end{equation}
which was seen to fit very well with that obtained in experiment.  

One notices first the striking fact that the upper right corner (i.e. the 13) 
elements in both matrices are particularly small and much smaller than 
the 12 elements (i.e. the Cabibbo angle) or the 23 elements.  For the quark 
case, the smallness of the 13 element is needed for explaining $b$ decays, 
while for the lepton case, we recall, it is needed to satisfy the bounds 
imposed by the CHOOZ \cite{Chooz} oscillation experiment, at least for
$m_3^2 > 10^{-3} {\rm eV}^2$.  It would thus be interesting to understand 
why this feature should emerge correctly from our calculations with the 
DSM scheme.  The answer turns out to be quite intriguing, giving the above 
feature as a consequence of the general differential geometric properties 
of space curves.  As already explained, the non-diagonal CKM matrix 
elements arise 
from loop-corrections which forces the vector $(x',y',z')$ to run along 
a trajectory on a sphere, as depicted in Figure \ref{runtraj} above.  
Recalling then from \cite{OurCKM} in detail how the state vectors of the 
three physical states of each fermion-type are defined and how the CKM 
matrix is constructed from these, it can be shown \cite{curves} that the 
12 and 23 elements of the CKM matrix are associated with the curvatures 
of the trajectory on the sphere while the element 13 is associated with 
its torsion.  It then follows that the 13 element is necessarily small 
compared with the 12 and 23 elements.

Secondly, we note that the 23 element is much larger in the lepton than 
in the quark CKM matrix.  This is physically important, or otherwise, as 
explained above, one would not be able to explain the large muon anomaly
observed in atmospheric neutrinos.  Within the scheme employed here, this 
enhancement of the 23 element for leptons over quarks can again be easily 
understood in differential geometrical terms.  Indeed, it can be shown 
\cite{curves} that the 23 element of a CKM matrix is associated with the 
so-called normal curvature of the trajectory which on a sphere is constant.
This element is thus roughly proportional to the separation on the trajectory 
between the locations of the two fermion-types to which it refers.  Now,
as can be seen in Figure \ref{runtraj}, the leptons have a much longer 
distance to run from $\tau$ to $\nu_3$ than the quarks from $t$ to $b$, so 
that it follows that $U_{\mu 3}$ is necessarily much larger than $V_{cb}$,
as is experimentally observed.

The fact that these important empirical features can be traced through some 
simple differential geometry directly back to the intrinsic properties of 
the Dualized Standard Model we consider to be a nontrivial and encouraging 
check both of the scheme itself and of our calculations.

So far, we have considered only the case (i) with $m_2^2 \sim 10^{-10}$ eV$^2$
corresponding to the so-called LWO solution to the solar neutrino problem.
What about the case (ii) with $m_2^2 \sim 10^{-5}$ eV$^2$ corresponding to
the so-called MSW solution?  This is not as easy for the present scheme 
to accommodate.  We recall that in the DSM scheme, the second generation 
acquires a mass only through `leakage' from the highest generation, and
this `leakage' is limited, as explained above, by the curvature of the
trajectory.  Given that the value of $m_2$ required by the MSW solution (ii)
is so much larger than that required by the LWO solution (i), the former
will require a much larger `leakage' and this is not easily available
on our trajectory.  Indeed, taking the same value of $\rho$ for neutrinos 
as for the other fermions as we have done above, one can easily find 
the maximum value for $M_2/M_3$ to be about 0.11, which for $m_2^2$ of order 
$10^{-5}$ eV$^2$ gives necessarily $m_3^2 > 7 \times 10^{-2} {\rm eV}^2$, 
which is some way above the range preferred by Kamiokande \cite{Kamioka}
and SuperKamiokande \cite{superk}.  If one relaxes the condition that
$\rho$ should be the same as for the other fermions, for which after all
there is as yet no theoretical justification, then one can obtain enough
`leakage' to move $m_3^2$ into the $10^{-2}-10^{-3} {\rm eV}^2$ range,
but only at the cost of a large $\rho > 5$ and a very large Dirac mass
$M_3 \sim {\rm TeV}$.  Besides, it requires further struggle to get the
mixing angles within the experimental bounds set by e.g. \cite{Giunkimno}
since the sizeable value for $U_{\mu 3}$ required by the atmospheric
neutrino data necessitates, in our present framework, also a sizeable
separation on the trajectory between the locations of the two highest 
generation neutrinos.  Indeed, in all the attempts we have made so far, it 
is only by choosing values as high as $\rho \sim 18, M_3 \sim 16 {\rm TeV}$
that we manage to get $m_3^2 \sim 10^{-2} {\rm eV}^2$ and $U_{\mu 3}
\sim 0.44$ just within the the 90\% limits set by \cite{Giunkimno} from
the Kamiokande data.  It thus seems that although one can still possibly 
obtain some fits at the cost of one more parameter $\rho$ than the 
case above for $m_2^2 \sim 10^{-10} {\rm eV}^2$, this scenario for 
$m_2^2 \sim 10^{-5}{\rm eV}^2$ is far less comfortably accommodated.  

It is clear also that the DSM scheme would have difficulty accommodating
neutrinos with masses of the order of several eV's as those wanted by 
astrophysicists for hot dark matter \cite{hotdark}, or the neutrinos 
possibly indicated by the LSND \cite{LSND} experiment.  One can, of 
course, introduce here by hand, as one does in other schemes, extra 
sterile neutrinos to foot the bill, but that would be against the spirit 
of the whole idea which is, perhaps over-ambitiously, to aim at an overall
explanation for the quark and lepton spectrum as we know it today.

However, in the case with $m_2^2 \sim 10^{-10} {\rm eV}^2$ with which the 
DSM scheme is most happy, one is able to predict with no free parameter 
all the mixing angles which appear to be consistent with what is known 
so far in experiment.  And these results are obtained with the same method 
as that applied before to calculate the quark CKM matrix using exactly 
the same values of the common parameters.  It is this possibility of a 
consistent treatment of the two related problems that we find most 
encouraging.

\vspace{.5cm}
\noindent{\large {\bf Acknowledgement}}\\

We are much indebted to Roger Phillips and Bill Scott for some helpful
conversations on the experimental situation in neutrino oscillations.
One of us (JB) acknowledges support from the Spanish Government on
contract no. CICYT AEN 97-1718, while another (JP) is grateful to the
Studientstiftung d.d. Volkes and the Burton Senior Scholarship of
Oriel College, Oxford for financial support.

\end{document}